\documentclass[10pt,aps,prl,twocolumn,superscriptaddress]{revtex4-2}
\usepackage{braket}
\usepackage{amsfonts, mathtools}
\usepackage{xcolor}
\definecolor{darkblue}{rgb}{0, 0, 0.5}
\usepackage[breaklinks=true,colorlinks,citecolor=darkblue,linkcolor=darkblue,urlcolor=darkblue]{hyperref}
\usepackage{orcidlink}

\newcommand{\norm}[1]{\left\lVert#1\right\rVert}
\newcommand{\var}[1]{\text{Var}(#1)}
\newcommand{\figlabel}[1]{Fig.~\ref{#1}}

\begin{document}

\title{Scrambling in the Charging of Quantum Batteries}
\author{Sebastián V. Romero$^{\orcidlink{0000-0002-4675-4452}}$}
\affiliation{Department of Physical Chemistry, University of the Basque Country UPV/EHU, Apartado 644, 48080 Bilbao, Spain}
\affiliation{Kipu Quantum GmbH, Greifswalderstrasse 212, 10405 Berlin, Germany}
\author{Yongcheng Ding$^{\orcidlink{0000-0002-6008-0001}}$}
\affiliation{Department of Physical Chemistry, University of the Basque Country UPV/EHU, Apartado 644, 48080 Bilbao, Spain}
\affiliation{Institute for Quantum Science and Technology, Department of Physics, Shanghai University, Shanghai 200444, China}
\author{Xi Chen$^{\orcidlink{0000-0003-4221-4288}}$}
\email{xi.chen@csic.es}
\affiliation{Instituto de Ciencia de Materiales de Madrid (CSIC), Cantoblanco, E-28049 Madrid, Spain}
\author{Yue Ban$^{\orcidlink{0000-0003-1764-4470}}$}
\email{yue.ban@csic.es}
\date{\today}

\begin{abstract}
  Exponentially fast scrambling of an initial state characterizes quantum chaotic systems. Given the importance of quickly populating higher energy levels from low-energy states in quantum battery charging protocols, this work investigates the role of quantum scrambling in quantum batteries and its effect on optimal power and charging times by means of the Sachdev-Ye-Kitaev model, a maximally-chaotic black hole physics model that has been recently proposed as a quantum battery. We adopt a bare representation with normalized bandwidths to suppress system energy dependence. To our knowledge, this is the first in-depth exploration of quantum scrambling in the context of quantum batteries. By analyzing the dynamics of out-of-time-order correlators, our findings indicate that quantum scrambling does not necessarily lead to faster charging, despite its potential for accelerating the process.
\end{abstract}%

\maketitle

\emph{Introduction.}---A quantum battery (QB) is a system that stores energy in identical quantum cells, which can then be extracted as work. After a QB is charged, it is desirable to store its energy for a sufficiently long period before extraction~\cite{hovhannisyan2013entanglement}. Since 2013~\cite{alicki2013entanglement}, entangled unitary operators acting on quantum cells, known as \emph{collective charging}~\cite{campaioli2017enhancing}, have been proposed to enhance work extraction in comparison with the unentangled controls performance, referred to as \emph{parallel charging}~\cite{binder2015quantacell}.  Several models have been put forth, with recent experimental realizations~\cite{campaioli2023colloquium} on superconductors, quantum dots, organic microcavities, and nuclear spins. In refs.~\cite{rosa2020ultra-stable,rossini2020quantum}, the authors explore the exactly solvable Sachdev–Ye–Kitaev (SYK) model as a QB. Leveraging strong nonlocal correlations, their findings reveal that this SYK-based mechanism leads to a highly stable charging protocol, demonstrating superextensive scaling of average power with system size. As a result, SYK QBs are capable of outperforming any classical counterpart, thereby providing a clear quantum advantage.

Optimizing battery charging is naturally a key focus in QBs. To achieve faster and more reliable charging processes in the near term, appropriate quantum control techniques are essential.
Here a critical challenge is accelerating the population of higher-excited states, where quantum chaos could play a pivotal role. 
Although the precise definition of quantum chaos remains debated~\cite{xu2020does,dowling2023scrambling}, it is widely regarded as one of the most suitable quantities to study scrambling in chaotic systems. This is typically explored through out-of-time-order correlators (OTOC)~\cite{larkin1969quasiclassical,garttner2017measuring,swingle2018unscrambling}, Loschmidt echo~\cite{peres1984stability,jalabert2001environment}, and the butterfly effect~\cite{Lieb1972finite}, among others.
\begin{figure}[!tb]
    \centering
    \includegraphics[width=\linewidth]{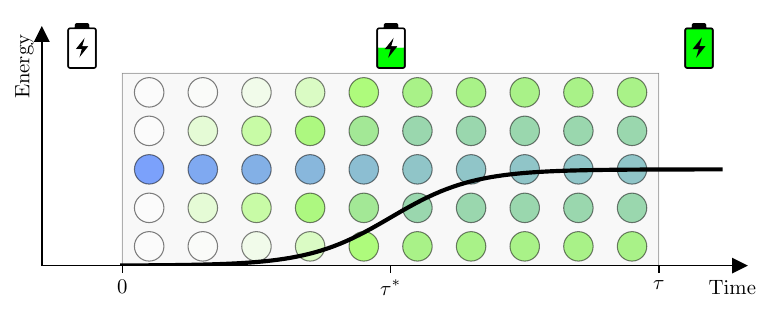}\vspace{-5mm}%
    \caption{Schematic diagram for charging process. Starting from the ground state of a battery Hamiltonian $\mathcal{H}_0$ (battery discharged) a charger $\mathcal{H}_1$ is plugged in within the time window $t\in[0,\tau]$ to populate higher excited states. This protocol starts in a separated state and ends up in a completely mixed state for sufficiently long charging times, being this feature a potential signature of quantum chaos, {represented by the dots in an initially-localized operator spreading fashion~\cite{swingle2018unscrambling}}.}\label{fig:fig1}
\end{figure}%

In this work, we examine the dynamical properties of scrambling in QB charging performance using OTOC and the butterfly effect. 
The all-to-all connectivity of the SYK model is known to achieve superextensive scaling. 
Additionally, a key feature of SYK models is their maximal chaos.
On one hand, quantum chaos can be used to scramble an initial state faster, which could become a potential resource for faster charging protocols (see~\figlabel{fig:fig1} for a schematic diagram). On the other hand, it is known that the amount of chaos decays with the system size~\cite{kobrin2021manybody,caceres2023outoftimeorder}, causing an increase in the Ehrenfest time~\cite{adamov2003loschmidt, lewis2019unifying, te}. This appears to conflict with the characteristic scaling of optimal charging time, which decays with the system size as $\tau^*\sim N^{-1/2}$~\cite{rossini2020quantum}. Since time depends on energy, which could introduce undesired bias in our study, we consider a regularized framework. In this approach, the bandwidths, defined as the difference between the lowest and largest eigenvalues of the charger Hamiltonian, are normalized to allow a fair comparison. This bare representation enable a more accurate analysis of how chaos spreads across the quantum battery and its dependence on the system size, without the influence of energy units.

\emph{QB charging under complex SYK.}---We define the system Hamiltonian as $\mathcal{H}(t) = \mathcal{H}_0 + \lambda(t)(\mathcal{H}_1-\mathcal{H}_0)$, where $\mathcal{H}_0 = \sum_{j=1}^N h_j$ is the battery Hamiltonian with $N$ quantum cells, and $h_j=\omega_0\sigma^y_j/2$. Here $\mathcal{H}_1$ is the charging Hamiltonian, and $\lambda(t)$ is a switching function that toggles between $\mathcal{H}_0$ and $\mathcal{H}_1$. We use a unit step function with $\lambda(t) = 1$ for $t\in[0,\tau]$ and $0$ otherwise. The charging Hamiltonian $\mathcal{H}_1$ is given by the complex SYK model~\cite{sachdev2015bekenstein}:
\begin{equation}
\label{eq:c-syk}
\mathcal{H}_1=\sum_{i,j,k,l=1}^N \mathcal{J}_{ijkl}c^\dagger_i c^\dagger_j c_k c_l,
\end{equation}
where $c^\dagger_i$ ($c_i$) creates (annihilates) a spinless fermion on site $i$ under the Jordan-Wigner transformation. The couplings $\mathcal{J}_{ijkl}$ are zero-mean, Gaussian-distributed complex random variables with variance $\var{\mathcal{J}_{ijkl}}=J^2/N^3$. They satisfy the conditions $\mathcal{J}_{ijkl}=\mathcal{J}^*_{klij}=-\mathcal{J}_{jikl}=-\mathcal{J}_{ijlk}$, where $J$ represents the energy units of $\mathcal{H}_1$. The charging Hamiltonian must satisfy $[\mathcal{H}_0, \mathcal{H}_1]\neq 0$ to effectively inject energy into the system. Hereinafter, all results presented are averaged over 1000, 500, and 200 different realizations of the complex SYK couplings $\mathcal{J}_{ijkl}$ for $N \in [4, 10]$, $N \in [11, 13]$, and $N \in [14, 17]$, respectively.

Let $\ket{\psi(\tau)}$ denote the state of the system after charging completion. The mean local energy injected is defined as $E_N(\tau)=\braket{\psi(\tau)|\mathcal{H}_0|\psi(\tau)}$, measured in units of $\omega_0$ (with $\hbar \equiv 1$). The average charging power is then given by $P_N(\tau)=E_N(\tau)/\tau$ (in units of $\omega_0 J$). This definition of power establishes a trade-off between maximizing the amount of energy stored and minimizing the time required for charging. Accordingly, the optimal charging time $\tau^*$ is defined as the time at which the maximum average power is achieved, that is, $P_N(\tau^*)=\max_\tau P_N(\tau)$.

\emph{Speedup of charging by means of chaos?}---Now we begin with a preliminary test of scrambling dynamics and system size dependence via OTOC using $\mathcal{H}_1$ under conventional approach, with $J=\omega_0=1$. Based on~\cite{garttner2017measuring,swingle2018unscrambling}, we define OTOC as
\begin{equation}\label{eq:otoc}
    F(t)\coloneqq1-|\braket{W^\dagger(t)V^\dagger W(t)V}_0|^2,
\end{equation}
where $V\coloneqq c_N+c^\dagger_N$ and $W\coloneqq c_{N-1}+c^\dagger_{N-1}$ are chosen as local operators, and $W(t)= e^{i\mathcal{H}_1t} W e^{-i\mathcal{H}_1t}$ with $\braket{\cdot}_t \coloneqq \braket{\psi(t)|\cdot|\psi(t)}$. This function measures how fast two initially commuting operators fail to commute. Since our choice of local unitary operators satisfies $[W,V]=0$, it follows that $F(0)=0$ and $0\le F(t)\le1$. The exponential growth at shorter times is used to quantify scrambling, where $F(t)\sim e^{\lambda_\text{L}t}$ with $\lambda_\text{L}$ as the quantum Lyapunov exponent. Inspired by Ref.~\cite{kobrin2021manybody}, we fit each OTOC curve to $a+be^{\lambda_\text{fit}t}$ within a range defined by $F_0\le F(t)\le F_1$, where the bounds are rough estimates of the dissipation time and a large enough time to capture the exponential growth for all curves, respectively. We set $F_0=0.02$ and $F_1=0.2$ for convenience. From here, we expand $\lambda_\text{fit}$ in terms of $1/N$ as $\lambda_\text{fit}(N)\coloneqq\lambda_0+\lambda_1/N+\lambda_2/N^2$. As shown in~\figlabel{fig:fig2}, the quantum Lyapunov exponent decreases with the system size, resulting in an increase in Ehrenfest times $t_\text{E}\sim\log(N)/\lambda_\text{L}$~\cite{lewis2019unifying}. This finding contradicts the intuitive expectation that scrambling might be a key source of faster charging protocols, characterized by the scaling $\tau^*\sim N^{-1/2}$ for complex SYK QBs~\cite{rossini2020quantum}. In addition, $F(\tau^*)$ values  decrease with the system size, even though $\tau^*$ does as well.
\begin{figure}[!tb]
    \centering
    \includegraphics[width=\linewidth]{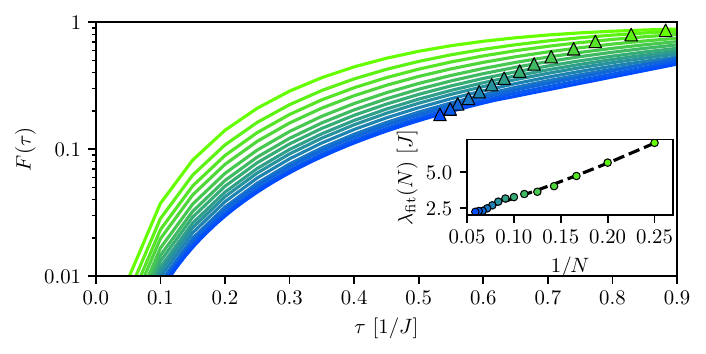}\vspace{-5mm}%
    \caption{OTOC dynamics of charging process setting $J=\omega_0=1$. From upper to lower curves, $N\in[4,17]$ in increasing order, where $F(\tau^*)$ values are marked. Inset includes the decaying fitted Lyapunov exponents with the inverse of $N$, which apparently contradicts the superextensive scaling behavior characteristic of complex SYK QBs.}\label{fig:fig2}
\end{figure}%

The scaling of the optimal charging time can be understood through the Fubini-Study distance~\cite{anandan1990geometry}, which measures the trajectory traced  by $\ket{\psi(t)}$ over the interval $t \in [0, \tau]$ along the curve $\mathcal{C}$ in the projective Hilbert space. This distance is given by $l(\mathcal{C}) = \Delta_\tau \mathcal{H}_1 \tau$, where $\Delta_\tau \mathcal{H}^\alpha = \frac{1}{\tau}\int_0^\tau \text{d}t \big[\braket{\mathcal{H}^2(t)}_t-\braket{\mathcal{H}(t)}_t^2 \big]^{\alpha/2}$ for $\alpha \in \{1,2\}$, representing the averaged uncertainty and variance, respectively. Since $\tau \sim 1/\Delta_\tau \mathcal{H}_1$ when $l(\mathcal{C})$ saturates, and based on the energy variances presented in Ref.~\cite{rossini2020quantum} and computed in Supplemental Material~\cite{sm}, this relation directly leads to the scaling law $\tau^* \sim N^{-1/2}$.

Based on these results, despite the fact that the amount of scrambling decreases with $N$, the optimal charging times also decrease. This indicates that the superextensive scaling does not come from how the charger scrambles. Instead, it may be related to the system energy dependence, which could introduce undesirable side effects and lead to a seemingly contradictory outcome.

\emph{Bandwidth regularization.}---To resolve the inconsistency in the previous conventional analysis, we first shift the spectrum of the charger $\mathcal{H}_1$ to zero. Then, we  regularize it by applying the mapping $\mathcal{H}_1\mapsto\mathcal{H}_1/\norm{\mathcal{H}_1}$, where $\norm{\cdot}=\mu(\cdot)$ denotes its norm, with $\mu(\cdot)$ being the largest singular value. This bandwidth-regularized charger eliminates spurious effects arising from energy dependence.
\begin{figure*}[!tb]
    \centering
    \includegraphics[width=\linewidth]{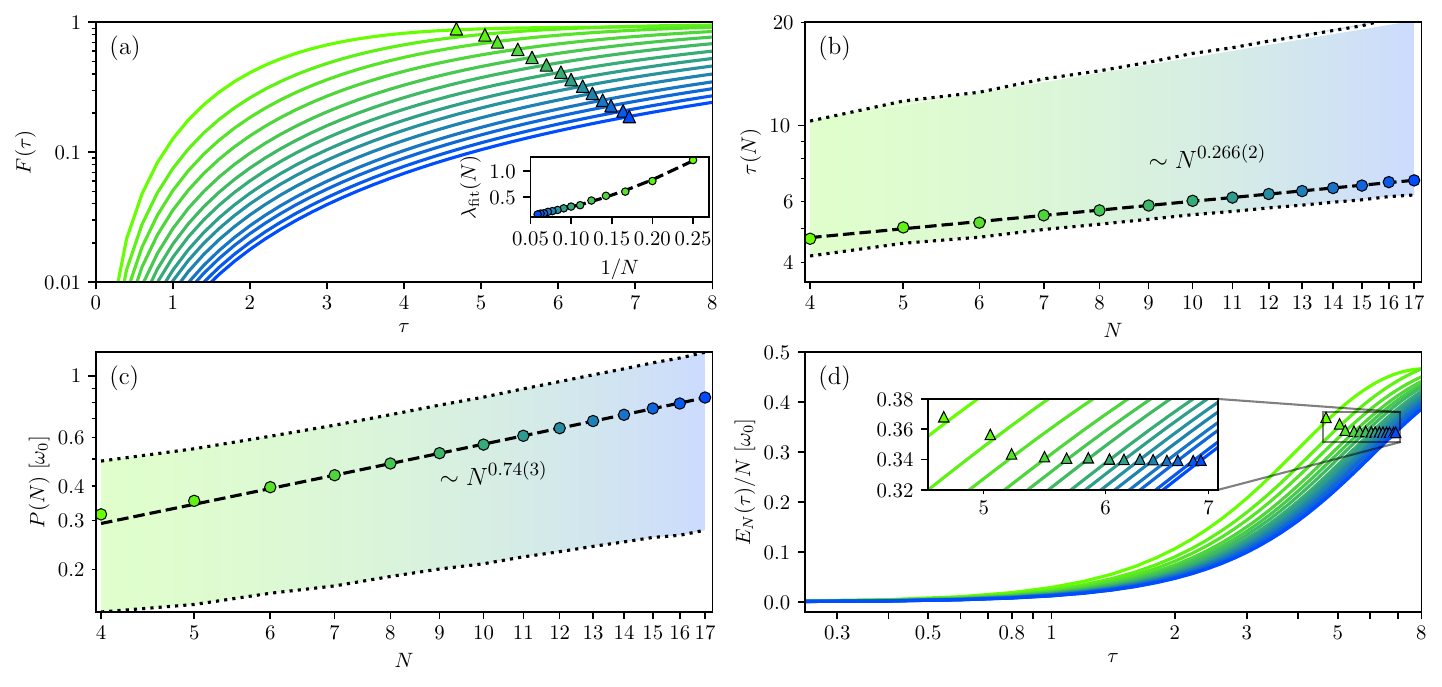}\vspace{-5mm}%
    \caption{Under bandwidth regularization: (a) OTOC dynamics. From upper to lower curves, $N\in[4,17]$ in increasing order, where $F(\tau^*)$ values are marked. Inset includes the decaying fitted Lyapunov exponents with the inverse of the system size, similar to the results obtained in~\figlabel{fig:fig2}. (b) Normalized mean local energy dynamics. From upper to lower curves, $N\in[4,17]$ in increasing order. Marks correspond to the optimal charging times $\tau^*$, which are reached around $1/3$ which is below the $1/2$ value expected for a completely mixed state (see~\figlabel{fig:fig4}). (c)-(d) Optimal charging times and powers, respectively, for $N\in[4,17]$ cells with their power-law fittings $a+bN^c$ in dashed lines, where the three data points corresponding to the smallest $N$ have been discarded from the fits. In dotted lines, charging times are bounded by $T_\text{(R)QSL}$ and powers by eq.~\eqref{eq:power_bounds}. After regularization, (c) shows how optimal charging times increase with $N$ under this representation and (d) how optimal powers still increase with $N$ due to its inherent superextensive scaling nature.}\label{fig:fig3}
\end{figure*}%

Before hurrying toward the corrected charging-speedup analysis, it is essential to point out that both optimal charging time and power are subject to upper and lower bounds. Specifically, the quantum speed limit (QSL)~\cite{chen2010transient,deffner2013energytime} provides a lower bound on the evolution time of a quantum system from an initial to a target state. It is given by $T_\text{QSL}\coloneqq\max \{ \mathcal{L}(\psi_0,\psi_\tau)/E,\mathcal{L}(\psi_0,\psi_\tau)/\Delta E \} $, where $\mathcal{L}(\psi_0,\psi_\tau)=\cos^{-1}|\braket{\psi(0)|\psi(\tau)}|$ is the Bures angle, $E= \frac{1}{\tau}\int_0^\tau\text{d}t\braket{\mathcal{H}}_t$ is the average energy, and $\Delta E= \Delta_\tau\mathcal{H}_1$. On the contrary, the reverse quantum speed limit (RQSL) sets an upper bound on the evolution time~\cite{mohan2021reverse}. It is defined as $T_\text{RQSL}\coloneqq \ell(\chi(\tau))/\Delta E$, where the reference section relative to the initial state reads $
\ket{\chi(t)}\coloneqq \frac{\braket{\psi(t)|\psi(0)}}{|\braket{\psi(t)|\psi(0)}|}\ket{\psi(t)}$,
and the geometric length is given $\ell(\chi(\tau))\coloneqq \int_0^\tau\text{d}t \sqrt{\braket{\dot{\chi}(t)|\dot{\chi}(t)}}$.
Thus, both speed limits establish lower and upper bounds on charging time, ensuring that $T_\text{QSL}\le\tau\le T_\text{RQSL}$. 

Similarly, following refs.~\cite{rossini2020quantum,mohan2021reverse} we can derive the corresponding upper and lower bounds on the average charging power $P_N(\tau)$ as follows:
\begin{equation}\label{eq:power_bounds}
 \frac{E_N(\tau)}{T_\text{RQSL}} \leq P_N(\tau) \leq 2\sqrt{\Delta_\tau \mathcal{H}_0^2 \Delta_\tau \mathcal{H}_1^2}.
\end{equation}
The variance $\Delta_\tau\mathcal{H}_0^2$ relates to the distance traveled in the Hilbert space, revealing the quantum nature along the charging process. Larger values of this variance correspond to shorter trajectories in the Hilbert space, transitioning from an initially pure to a highly entangled state, thereby increasing the charging power. Moreover, $\Delta_\tau\mathcal{H}_1^2$ directly reflects the charging speed, where larger values naturally lead to faster charging processes~\cite{sm}.
\begin{figure*}[!tb]
    \centering
    \includegraphics[width=.5\linewidth]{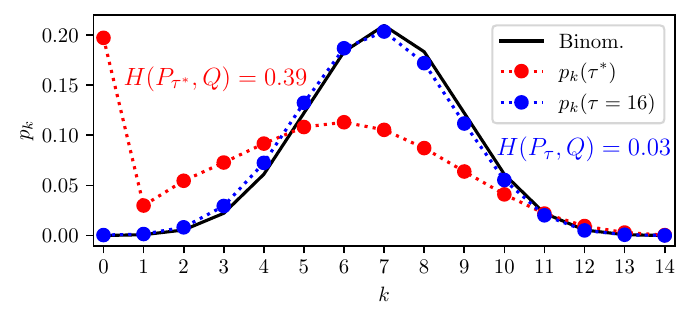}%
    \includegraphics[width=.5\linewidth]{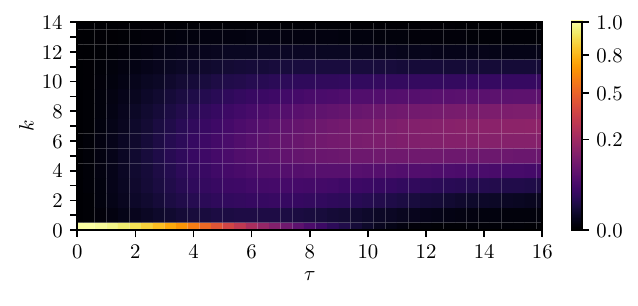}\vspace{-5mm}
    \caption{Left: populations distribution at optimal regularized charging time (red), $p_k(\tau^*\sim6)$, and large charging time (blue), $\tau=16$; for $N=14$ against a binomial distribution (black). The corresponding Hellinger distances are attached. Right: dynamics of the populations $p_k(\tau)$ for the same system, showing that at $\tau^*$ the completely mixed state region is not reached as expected, meaning that completely mixing our initial state is not required for optimally charging a QB. Nonregularized chargers show similar results.}\label{fig:fig4}
\end{figure*}%
\begin{figure*}[!t]
    \centering\vspace{-3mm}
    \includegraphics[width=\linewidth]{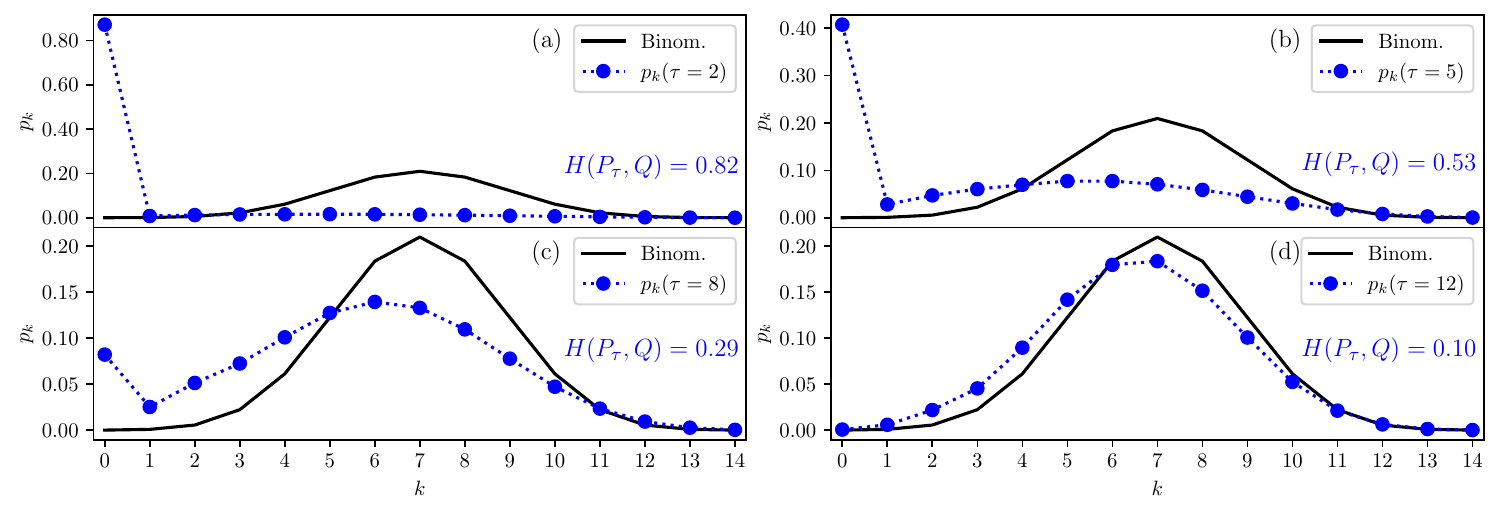}\vspace{-5mm}%
    \caption{(a)-(d) Populations distribution at different charging times $\tau\in\{2,5,8,12\}$ (blue) for a regularized charger with $N=14$ cells against a binonmial distribution (black). The corresponding Hellinger distances are attached.}\label{fig:fig5}
\end{figure*}%

Contrary to the findings in~\figlabel{fig:fig2}, within the regularized framework, while the quantum Lyapunov exponents decrease [\figlabel{fig:fig3}a], the optimal charging times now exhibit an increase with the system size, as demonstrated in~\figlabel{fig:fig3}b. These results stem from the removal of energy units, making both representations interchangeable by simply multiplying quantities by the $\mathcal{H}_1$ bandwidth accordingly. In this bare representation, we observe that, under our definition of quantum chaos as scrambling [eq.~\eqref{eq:otoc}] and with the charger Hamiltonian $\mathcal{H}_1$, faster quantum scrambling signatures are intrinsically associated with faster charging processes. Furthermore, our results show that the bandwidth of the charging Hamiltonian, governed by $J$, is the key factor to accelerate the charging, which in turn translates into faster scrambling. The optimal powers, shown in~\figlabel{fig:fig3}c, highlight their growth with $N$ in the bare representation, reinforcing the advantage of employing complex SYK QB models. It is well-known that in the large $N$ limit, the bandwidth of the SYK model scales as $\mathcal{O}[N]$, feature of interacting fermion models. Therefore, we would expect in this limit that the optimal charging times and powers scale as $\sim N^{1/2}$ in figures~\ref{fig:fig3}b-c, not obtained because of finite size effects. However, the fitted data is in accordance with the expected extensive scaling in the energy injected and the Fubini-Study distance discussion previously mentioned~\cite{sm}.

In~\figlabel{fig:fig3}d, the normalized $E_N(\tau)$ is plotted as a function of charging time for different $N$, with their corresponding optimal charging times marked. The region where the populations $p_k$ of the $k$th energy level stabilize corresponds to a binomial distribution $p_k\sim\binom{N}{k}/2^N$, which aligns with the expected results for a maximally mixed state $\rho=I/2^N$. This distribution yields $E_N(\tau)\sim N\omega_0/2$, which is notably larger than $E_N(\tau^*)$. However, this region is reached considerably after optimal charging times, which is explained by the way $\tau^*$ is defined~\cite{sm}. To show that, the spectrum degeneracy allows studying neatly how mixed the state is along the evolution, thus the effect of an all-to-all connected complex SYK over an initially completely separated state. By means of the Hellinger distance~\cite{hellinger1909neue}, it is possible to study the distance between the populations dynamics $p_k(\tau)$ for a fixed $\tau$ and a binomial distribution, which characterizes the population distribution for a completely mixed state. The corresponding Hellinger distance is computed as
\begin{equation}\label{eq:hellinger}
    H^2(P_\tau,Q) = 1 - \frac{1}{2^{N/2}}\sum_{k=0}^N \sqrt{p_k(\tau)\binom{N}{k}},
\end{equation}
with $P_\tau\coloneqq\{p_k(\tau)\}_{k=0}^N$, $Q\coloneqq\{\binom{N}{k}/2^N\}_{k=0}^N$ and $0\le H^2(P_\tau,Q)\le1$.

In~\figlabel{fig:fig4}, for a bandwidth-regularized complex SYK QB with $N=14$ cells, we plot the binomial and the population distributions at two different charging time regimes, at optimal charging time $\tau^*\sim 6$ [\figlabel{fig:fig3}b] and a considerably larger time $\tau=16$, where their corresponding Hellinger distances using eq.~$\eqref{eq:hellinger}$ are attached. Moreover, population dynamics are presented as a heatmap to visualize how different energy levels get populated against time. Both plots together showcase that the optimal charging times are reached before completely mixing our initial state. This is confirmed by the large value of the Hellinger distance obtained for $\tau^*$. Nonetheless, as the state keeps evolving, higher-state energy levels get more populated to the detriment of the lower ones and its corresponding populations approaches a binomial distribution. In~\figlabel{fig:fig5}, we depict more thoroughly the dynamics of the populations, showing how higher-energy states get populated along the evolution under a regularized complex SYK charger by computing the Hellinger distance between the populations and the binomial distribution at different charging times.

\emph{Scrambling from a butterfly effect perspective.}---From the OTOC form, we find that the only time-dependent operator involved—and thus the only one causing the OTOC to evolve—stems from the Heisenberg picture of $W$. This operator encapsulates how the system scrambles, starting from an initially localized operator that spreads across the system over time~\cite{swingle2018unscrambling}. This behavior is evident from expanding
\begin{equation}\label{eq:bch_formula}
    W(t)=\sum_{k=0}^{\infty} \frac{(it)^k}{k!} [\mathcal{H}_1, W]_k,
\end{equation}
which is a norm-convergent power series with $[\mathcal{H}_1, W]_0=W$ and $[\mathcal{H}_1, W]_k=[\mathcal{H}_1,[\mathcal{H}_1, W]_{k-1}]$. Starting from $W$ ($k=0$), this nested-commutator expansion shows that the effect of $\mathcal{H}_1$ on all fermions becomes more pronounced as time increases, with higher-order terms of the expansion becoming more relevant.

After removing energy units, we can analyze the system size dependence through the nested-commutators in eq.~\eqref{eq:bch_formula}, which uniquely contain the information on how the system behaves. In~\figlabel{fig:fig6}, we compute the norm of the first six nontrivial nested-commutators for various system sizes. For each order, their corresponding decays can be linked to the OTOC decay results obtained in~\figlabel{fig:fig3}a. Consequently, to achieve optimal charging times, larger charging protocols are required to compensate for the reduced scrambling with increasing $N$.
\begin{figure}[!t]
    \centering
    \includegraphics[width=\linewidth]{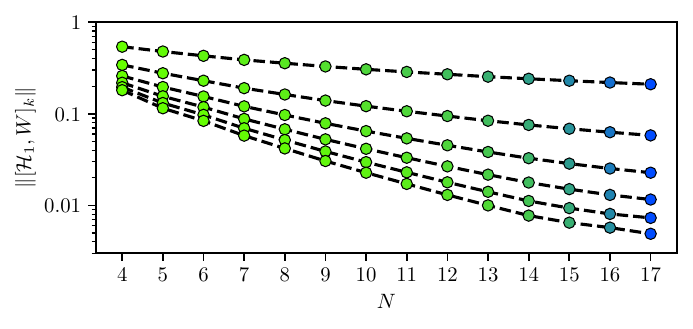}\vspace{-5mm}%
    \caption{From top to bottom curves in increasing order, norms of the first six nontrivial nested-commutators terms ($k=1, 2, ..., 6$) of eq.~\eqref{eq:bch_formula} against the system size. Their decay indicates that the scrambling spreading gets worse, thus has to be compensated by larger times.}\label{fig:fig6}
\end{figure}%

\emph{Conclusion.}---To advance QB charging protocols and accelerate the population of higher-excited states, we explore the use of scrambling as a crucial tool. This involves scrambling an initially unentangled state more rapidly. Although complex SYK QBs exhibit superextensive scaling, where optimal charging times decrease with the number of quantum cells, our OTOC analysis uncovers a counterintuitive result: quantum Lyapunov exponents decay with the system size. Our results, and our particular definition of quantum chaos, suggest that the role of quantum scrambling is more intricately tied to the amount of energy available in the charging Hamiltonian, rather than the scrambling dynamics itself. 
In our regularized framework, where energy units are normalized out, we observe that while quantum Lyapunov exponents continue to decrease with system size, optimal charging times actually increase. This shift underscores the importance of energy variance, which plays a crucial role in enabling the occupation of higher-excited states in the QB. To further understand the interplay between scrambling and system size, we also consider the butterfly effect. Our results indicate that the ability to populate higher-excited states in a QB is more dependent on the energy available at the charger and its connectivity rather than its scrambling behavior. Further exploration of how QB performance varies with these factors across different system sizes and graph-based connectivities among fermions~\cite{kim2022operator, divi2024sykchargingadvantagerandom}, as well as the impact of temperature, remains as future work.

\emph{Acknowledgments.}---We thank Dario Rosa, Juan José García-Ripoll and Juan Santos-Suárez for fruitful discussions. This work is supported by the Basque Government through Grant No. IT1470-22, the project grant PID2021-126273NB-I00 funded by MCIN/AEI/10.13039/501100011033 and by “ERDF A way of making Europe” and “ERDF Invest in your Future”, Nanoscale NMR and complex systems (Grant Nos. PID2021-126694NB-C21 and PID2021-126694NA-C22), the Spanish Ministry of Economic Affairs and Digital Transformation through the QUANTUM ENIA project call-Quantum Spain project, the project OpenSuperQ+100 (101113946) of the EU Flagship on Quantum Technologies. National Natural Science Foundation of China (Grants No. 12075145 and 12211540002), STCSM (Grant No. 2019SHZDZX01-ZX04) and the Innovation Program for Quantum Science and Technology (Grant No. 2021ZD0302302). Y.B. acknowledges the Ramón y Cajal (RYC2023-042699-I) research fellowship. Y.D. thanks the European Commission for a Marie Curie PF grant (No. 101204580 FELQO).

\bibliography{bibfile}

\begin{thebibliography}{29}%
\makeatletter
\providecommand \@ifxundefined [1]{%
 \@ifx{#1\undefined}
}%
\providecommand \@ifnum [1]{%
 \ifnum #1\expandafter \@firstoftwo
 \else \expandafter \@secondoftwo
 \fi
}%
\providecommand \@ifx [1]{%
 \ifx #1\expandafter \@firstoftwo
 \else \expandafter \@secondoftwo
 \fi
}%
\providecommand \natexlab [1]{#1}%
\providecommand \enquote  [1]{``#1''}%
\providecommand \bibnamefont  [1]{#1}%
\providecommand \bibfnamefont [1]{#1}%
\providecommand \citenamefont [1]{#1}%
\providecommand \href@noop [0]{\@secondoftwo}%
\providecommand \href [0]{\begingroup \@sanitize@url \@href}%
\providecommand \@href[1]{\@@startlink{#1}\@@href}%
\providecommand \@@href[1]{\endgroup#1\@@endlink}%
\providecommand \@sanitize@url [0]{\catcode `\\12\catcode `\$12\catcode `\&12\catcode `\#12\catcode `\^12\catcode `\_12\catcode `\%12\relax}%
\providecommand \@@startlink[1]{}%
\providecommand \@@endlink[0]{}%
\providecommand \url  [0]{\begingroup\@sanitize@url \@url }%
\providecommand \@url [1]{\endgroup\@href {#1}{\urlprefix }}%
\providecommand \urlprefix  [0]{URL }%
\providecommand \Eprint [0]{\href }%
\providecommand \doibase [0]{https://doi.org/}%
\providecommand \selectlanguage [0]{\@gobble}%
\providecommand \bibinfo  [0]{\@secondoftwo}%
\providecommand \bibfield  [0]{\@secondoftwo}%
\providecommand \translation [1]{[#1]}%
\providecommand \BibitemOpen [0]{}%
\providecommand \bibitemStop [0]{}%
\providecommand \bibitemNoStop [0]{.\EOS\space}%
\providecommand \EOS [0]{\spacefactor3000\relax}%
\providecommand \BibitemShut  [1]{\csname bibitem#1\endcsname}%
\let\auto@bib@innerbib\@empty
\bibitem [{\citenamefont {Hovhannisyan}\ \emph {et~al.}(2013)\citenamefont {Hovhannisyan}, \citenamefont {Perarnau-Llobet}, \citenamefont {Huber},\ and\ \citenamefont {Ac\'{\i}n}}]{hovhannisyan2013entanglement}%
  \BibitemOpen
  \bibfield  {author} {\bibinfo {author} {\bibfnamefont {K.~V.}\ \bibnamefont {Hovhannisyan}}, \bibinfo {author} {\bibfnamefont {M.}~\bibnamefont {Perarnau-Llobet}}, \bibinfo {author} {\bibfnamefont {M.}~\bibnamefont {Huber}},\ and\ \bibinfo {author} {\bibfnamefont {A.}~\bibnamefont {Ac\'{\i}n}},\ }\href {https://doi.org/10.1103/PhysRevLett.111.240401} {\bibfield  {journal} {\bibinfo  {journal} {Phys. Rev. Lett.}\ }\textbf {\bibinfo {volume} {111}},\ \bibinfo {pages} {240401} (\bibinfo {year} {2013})}\BibitemShut {NoStop}%
\bibitem [{\citenamefont {Alicki}\ and\ \citenamefont {Fannes}(2013)}]{alicki2013entanglement}%
  \BibitemOpen
  \bibfield  {author} {\bibinfo {author} {\bibfnamefont {R.}~\bibnamefont {Alicki}}\ and\ \bibinfo {author} {\bibfnamefont {M.}~\bibnamefont {Fannes}},\ }\href {https://doi.org/10.1103/PhysRevE.87.042123} {\bibfield  {journal} {\bibinfo  {journal} {Phys. Rev. E}\ }\textbf {\bibinfo {volume} {87}},\ \bibinfo {pages} {042123} (\bibinfo {year} {2013})}\BibitemShut {NoStop}%
\bibitem [{\citenamefont {Campaioli}\ \emph {et~al.}(2017)\citenamefont {Campaioli}, \citenamefont {Pollock}, \citenamefont {Binder}, \citenamefont {C\'eleri}, \citenamefont {Goold}, \citenamefont {Vinjanampathy},\ and\ \citenamefont {Modi}}]{campaioli2017enhancing}%
  \BibitemOpen
  \bibfield  {author} {\bibinfo {author} {\bibfnamefont {F.}~\bibnamefont {Campaioli}}, \bibinfo {author} {\bibfnamefont {F.~A.}\ \bibnamefont {Pollock}}, \bibinfo {author} {\bibfnamefont {F.~C.}\ \bibnamefont {Binder}}, \bibinfo {author} {\bibfnamefont {L.}~\bibnamefont {C\'eleri}}, \bibinfo {author} {\bibfnamefont {J.}~\bibnamefont {Goold}}, \bibinfo {author} {\bibfnamefont {S.}~\bibnamefont {Vinjanampathy}},\ and\ \bibinfo {author} {\bibfnamefont {K.}~\bibnamefont {Modi}},\ }\href {https://doi.org/10.1103/PhysRevLett.118.150601} {\bibfield  {journal} {\bibinfo  {journal} {Phys. Rev. Lett.}\ }\textbf {\bibinfo {volume} {118}},\ \bibinfo {pages} {150601} (\bibinfo {year} {2017})}\BibitemShut {NoStop}%
\bibitem [{\citenamefont {Binder}\ \emph {et~al.}(2015)\citenamefont {Binder}, \citenamefont {Vinjanampathy}, \citenamefont {Modi},\ and\ \citenamefont {Goold}}]{binder2015quantacell}%
  \BibitemOpen
  \bibfield  {author} {\bibinfo {author} {\bibfnamefont {F.~C.}\ \bibnamefont {Binder}}, \bibinfo {author} {\bibfnamefont {S.}~\bibnamefont {Vinjanampathy}}, \bibinfo {author} {\bibfnamefont {K.}~\bibnamefont {Modi}},\ and\ \bibinfo {author} {\bibfnamefont {J.}~\bibnamefont {Goold}},\ }\href {https://doi.org/10.1088/1367-2630/17/7/075015} {\bibfield  {journal} {\bibinfo  {journal} {New Journal of Physics}\ }\textbf {\bibinfo {volume} {17}},\ \bibinfo {pages} {075015} (\bibinfo {year} {2015})}\BibitemShut {NoStop}%
\bibitem [{\citenamefont {Campaioli}\ \emph {et~al.}(2024)\citenamefont {Campaioli}, \citenamefont {Gherardini}, \citenamefont {Quach}, \citenamefont {Polini},\ and\ \citenamefont {Andolina}}]{campaioli2023colloquium}%
  \BibitemOpen
  \bibfield  {author} {\bibinfo {author} {\bibfnamefont {F.}~\bibnamefont {Campaioli}}, \bibinfo {author} {\bibfnamefont {S.}~\bibnamefont {Gherardini}}, \bibinfo {author} {\bibfnamefont {J.~Q.}\ \bibnamefont {Quach}}, \bibinfo {author} {\bibfnamefont {M.}~\bibnamefont {Polini}},\ and\ \bibinfo {author} {\bibfnamefont {G.~M.}\ \bibnamefont {Andolina}},\ }\href {https://doi.org/10.1103/RevModPhys.96.031001} {\bibfield  {journal} {\bibinfo  {journal} {Rev. Mod. Phys.}\ }\textbf {\bibinfo {volume} {96}},\ \bibinfo {pages} {031001} (\bibinfo {year} {2024})}\BibitemShut {NoStop}%
\bibitem [{\citenamefont {Rosa}\ \emph {et~al.}(2020)\citenamefont {Rosa}, \citenamefont {Rossini}, \citenamefont {Andolina}, \citenamefont {Polini},\ and\ \citenamefont {Carrega}}]{rosa2020ultra-stable}%
  \BibitemOpen
  \bibfield  {author} {\bibinfo {author} {\bibfnamefont {D.}~\bibnamefont {Rosa}}, \bibinfo {author} {\bibfnamefont {D.}~\bibnamefont {Rossini}}, \bibinfo {author} {\bibfnamefont {G.~M.}\ \bibnamefont {Andolina}}, \bibinfo {author} {\bibfnamefont {M.}~\bibnamefont {Polini}},\ and\ \bibinfo {author} {\bibfnamefont {M.}~\bibnamefont {Carrega}},\ }\href {https://doi.org/10.1007/JHEP11(2020)067} {\bibfield  {journal} {\bibinfo  {journal} {Journal of High Energy Physics}\ }\textbf {\bibinfo {volume} {2020}},\ \bibinfo {pages} {67} (\bibinfo {year} {2020})}\BibitemShut {NoStop}%
\bibitem [{\citenamefont {Rossini}\ \emph {et~al.}(2020)\citenamefont {Rossini}, \citenamefont {Andolina}, \citenamefont {Rosa}, \citenamefont {Carrega},\ and\ \citenamefont {Polini}}]{rossini2020quantum}%
  \BibitemOpen
  \bibfield  {author} {\bibinfo {author} {\bibfnamefont {D.}~\bibnamefont {Rossini}}, \bibinfo {author} {\bibfnamefont {G.~M.}\ \bibnamefont {Andolina}}, \bibinfo {author} {\bibfnamefont {D.}~\bibnamefont {Rosa}}, \bibinfo {author} {\bibfnamefont {M.}~\bibnamefont {Carrega}},\ and\ \bibinfo {author} {\bibfnamefont {M.}~\bibnamefont {Polini}},\ }\href {https://doi.org/10.1103/PhysRevLett.125.236402} {\bibfield  {journal} {\bibinfo  {journal} {Phys. Rev. Lett.}\ }\textbf {\bibinfo {volume} {125}},\ \bibinfo {pages} {236402} (\bibinfo {year} {2020})}\BibitemShut {NoStop}%
\bibitem [{\citenamefont {Xu}\ \emph {et~al.}(2020)\citenamefont {Xu}, \citenamefont {Scaffidi},\ and\ \citenamefont {Cao}}]{xu2020does}%
  \BibitemOpen
  \bibfield  {author} {\bibinfo {author} {\bibfnamefont {T.}~\bibnamefont {Xu}}, \bibinfo {author} {\bibfnamefont {T.}~\bibnamefont {Scaffidi}},\ and\ \bibinfo {author} {\bibfnamefont {X.}~\bibnamefont {Cao}},\ }\href {https://doi.org/10.1103/PhysRevLett.124.140602} {\bibfield  {journal} {\bibinfo  {journal} {Phys. Rev. Lett.}\ }\textbf {\bibinfo {volume} {124}},\ \bibinfo {pages} {140602} (\bibinfo {year} {2020})}\BibitemShut {NoStop}%
\bibitem [{\citenamefont {Dowling}\ \emph {et~al.}(2023)\citenamefont {Dowling}, \citenamefont {Kos},\ and\ \citenamefont {Modi}}]{dowling2023scrambling}%
  \BibitemOpen
  \bibfield  {author} {\bibinfo {author} {\bibfnamefont {N.}~\bibnamefont {Dowling}}, \bibinfo {author} {\bibfnamefont {P.}~\bibnamefont {Kos}},\ and\ \bibinfo {author} {\bibfnamefont {K.}~\bibnamefont {Modi}},\ }\href {https://doi.org/10.1103/PhysRevLett.131.180403} {\bibfield  {journal} {\bibinfo  {journal} {Phys. Rev. Lett.}\ }\textbf {\bibinfo {volume} {131}},\ \bibinfo {pages} {180403} (\bibinfo {year} {2023})}\BibitemShut {NoStop}%
\bibitem [{\citenamefont {Larkin}\ and\ \citenamefont {Ovchinnikov}(1969)}]{larkin1969quasiclassical}%
  \BibitemOpen
  \bibfield  {author} {\bibinfo {author} {\bibfnamefont {A.~I.}\ \bibnamefont {Larkin}}\ and\ \bibinfo {author} {\bibfnamefont {Y.~N.}\ \bibnamefont {Ovchinnikov}},\ }\href {http://jetp.ras.ru/cgi-bin/e/index/e/28/6/p1200?a=list} {\bibfield  {journal} {\bibinfo  {journal} {Journal of Experimental and Theoretical Physics}\ }\textbf {\bibinfo {volume} {28}},\ \bibinfo {pages} {1200} (\bibinfo {year} {1969})}\BibitemShut {NoStop}%
\bibitem [{\citenamefont {Gärttner}\ \emph {et~al.}(2017)\citenamefont {Gärttner}, \citenamefont {Bohnet}, \citenamefont {Safavi-Naini}, \citenamefont {Wall}, \citenamefont {Bollinger},\ and\ \citenamefont {Rey}}]{garttner2017measuring}%
  \BibitemOpen
  \bibfield  {author} {\bibinfo {author} {\bibfnamefont {M.}~\bibnamefont {Gärttner}}, \bibinfo {author} {\bibfnamefont {J.~G.}\ \bibnamefont {Bohnet}}, \bibinfo {author} {\bibfnamefont {A.}~\bibnamefont {Safavi-Naini}}, \bibinfo {author} {\bibfnamefont {M.~L.}\ \bibnamefont {Wall}}, \bibinfo {author} {\bibfnamefont {J.~J.}\ \bibnamefont {Bollinger}},\ and\ \bibinfo {author} {\bibfnamefont {A.~M.}\ \bibnamefont {Rey}},\ }\href {https://doi.org/10.1038/nphys4119} {\bibfield  {journal} {\bibinfo  {journal} {Nature Physics}\ }\textbf {\bibinfo {volume} {13}},\ \bibinfo {pages} {781–786} (\bibinfo {year} {2017})}\BibitemShut {NoStop}%
\bibitem [{\citenamefont {Swingle}(2018)}]{swingle2018unscrambling}%
  \BibitemOpen
  \bibfield  {author} {\bibinfo {author} {\bibfnamefont {B.}~\bibnamefont {Swingle}},\ }\href {https://doi.org/10.1038/s41567-018-0295-5} {\bibfield  {journal} {\bibinfo  {journal} {Nature Physics}\ }\textbf {\bibinfo {volume} {14}},\ \bibinfo {pages} {988–990} (\bibinfo {year} {2018})}\BibitemShut {NoStop}%
\bibitem [{\citenamefont {Peres}(1984)}]{peres1984stability}%
  \BibitemOpen
  \bibfield  {author} {\bibinfo {author} {\bibfnamefont {A.}~\bibnamefont {Peres}},\ }\href {https://doi.org/10.1103/PhysRevA.30.1610} {\bibfield  {journal} {\bibinfo  {journal} {Phys. Rev. A}\ }\textbf {\bibinfo {volume} {30}},\ \bibinfo {pages} {1610} (\bibinfo {year} {1984})}\BibitemShut {NoStop}%
\bibitem [{\citenamefont {Jalabert}\ and\ \citenamefont {Pastawski}(2001)}]{jalabert2001environment}%
  \BibitemOpen
  \bibfield  {author} {\bibinfo {author} {\bibfnamefont {R.~A.}\ \bibnamefont {Jalabert}}\ and\ \bibinfo {author} {\bibfnamefont {H.~M.}\ \bibnamefont {Pastawski}},\ }\href {https://doi.org/10.1103/PhysRevLett.86.2490} {\bibfield  {journal} {\bibinfo  {journal} {Phys. Rev. Lett.}\ }\textbf {\bibinfo {volume} {86}},\ \bibinfo {pages} {2490} (\bibinfo {year} {2001})}\BibitemShut {NoStop}%
\bibitem [{\citenamefont {Lieb}\ and\ \citenamefont {Robinson}(1972)}]{Lieb1972finite}%
  \BibitemOpen
  \bibfield  {author} {\bibinfo {author} {\bibfnamefont {E.~H.}\ \bibnamefont {Lieb}}\ and\ \bibinfo {author} {\bibfnamefont {D.~W.}\ \bibnamefont {Robinson}},\ }\href {https://doi.org/10.1007/BF01645779} {\bibfield  {journal} {\bibinfo  {journal} {Communications in Mathematical Physics}\ }\textbf {\bibinfo {volume} {28}},\ \bibinfo {pages} {251–257} (\bibinfo {year} {1972})}\BibitemShut {NoStop}%
\bibitem [{\citenamefont {Kobrin}\ \emph {et~al.}(2021)\citenamefont {Kobrin}, \citenamefont {Yang}, \citenamefont {Kahanamoku-Meyer}, \citenamefont {Olund}, \citenamefont {Moore}, \citenamefont {Stanford},\ and\ \citenamefont {Yao}}]{kobrin2021manybody}%
  \BibitemOpen
  \bibfield  {author} {\bibinfo {author} {\bibfnamefont {B.}~\bibnamefont {Kobrin}}, \bibinfo {author} {\bibfnamefont {Z.}~\bibnamefont {Yang}}, \bibinfo {author} {\bibfnamefont {G.~D.}\ \bibnamefont {Kahanamoku-Meyer}}, \bibinfo {author} {\bibfnamefont {C.~T.}\ \bibnamefont {Olund}}, \bibinfo {author} {\bibfnamefont {J.~E.}\ \bibnamefont {Moore}}, \bibinfo {author} {\bibfnamefont {D.}~\bibnamefont {Stanford}},\ and\ \bibinfo {author} {\bibfnamefont {N.~Y.}\ \bibnamefont {Yao}},\ }\href {https://doi.org/10.1103/PhysRevLett.126.030602} {\bibfield  {journal} {\bibinfo  {journal} {Phys. Rev. Lett.}\ }\textbf {\bibinfo {volume} {126}},\ \bibinfo {pages} {030602} (\bibinfo {year} {2021})}\BibitemShut {NoStop}%
\bibitem [{\citenamefont {Cáceres}\ \emph {et~al.}(2023)\citenamefont {Cáceres}, \citenamefont {Guglielmo}, \citenamefont {Kent},\ and\ \citenamefont {Misobuchi}}]{caceres2023outoftimeorder}%
  \BibitemOpen
  \bibfield  {author} {\bibinfo {author} {\bibfnamefont {E.}~\bibnamefont {Cáceres}}, \bibinfo {author} {\bibfnamefont {T.}~\bibnamefont {Guglielmo}}, \bibinfo {author} {\bibfnamefont {B.}~\bibnamefont {Kent}},\ and\ \bibinfo {author} {\bibfnamefont {A.}~\bibnamefont {Misobuchi}},\ }\href {https://doi.org/10.1007/JHEP11(2023)088} {\bibfield  {journal} {\bibinfo  {journal} {Journal of High Energy Physics}\ }\textbf {\bibinfo {volume} {2023}},\ \bibinfo {pages} {88} (\bibinfo {year} {2023})}\BibitemShut {NoStop}%
\bibitem [{\citenamefont {Adamov}\ \emph {et~al.}(2003)\citenamefont {Adamov}, \citenamefont {Gornyi},\ and\ \citenamefont {Mirlin}}]{adamov2003loschmidt}%
  \BibitemOpen
  \bibfield  {author} {\bibinfo {author} {\bibfnamefont {Y.}~\bibnamefont {Adamov}}, \bibinfo {author} {\bibfnamefont {I.~V.}\ \bibnamefont {Gornyi}},\ and\ \bibinfo {author} {\bibfnamefont {A.~D.}\ \bibnamefont {Mirlin}},\ }\href {https://doi.org/10.1103/PhysRevE.67.056217} {\bibfield  {journal} {\bibinfo  {journal} {Phys. Rev. E}\ }\textbf {\bibinfo {volume} {67}},\ \bibinfo {pages} {056217} (\bibinfo {year} {2003})}\BibitemShut {NoStop}%
\bibitem [{\citenamefont {Lewis-Swan}\ \emph {et~al.}(2019)\citenamefont {Lewis-Swan}, \citenamefont {Safavi-Naini}, \citenamefont {Bollinger},\ and\ \citenamefont {Rey}}]{lewis2019unifying}%
  \BibitemOpen
  \bibfield  {author} {\bibinfo {author} {\bibfnamefont {R.~J.}\ \bibnamefont {Lewis-Swan}}, \bibinfo {author} {\bibfnamefont {A.}~\bibnamefont {Safavi-Naini}}, \bibinfo {author} {\bibfnamefont {J.~J.}\ \bibnamefont {Bollinger}},\ and\ \bibinfo {author} {\bibfnamefont {A.~M.}\ \bibnamefont {Rey}},\ }\href {https://doi.org/10.1038/s41467-019-09436-y} {\bibfield  {journal} {\bibinfo  {journal} {Nature Communications}\ }\textbf {\bibinfo {volume} {10}},\ \bibinfo {pages} {1581} (\bibinfo {year} {2019})}\BibitemShut {NoStop}%
\bibitem [{te()}]{te}%
  \BibitemOpen
  \href@noop {} {}\bibinfo {note} {Ehrenfest time is defined as the crossing point between exponential to power-law scaling of scrambling.}\BibitemShut {Stop}%
\bibitem [{\citenamefont {Sachdev}(2015)}]{sachdev2015bekenstein}%
  \BibitemOpen
  \bibfield  {author} {\bibinfo {author} {\bibfnamefont {S.}~\bibnamefont {Sachdev}},\ }\href {https://doi.org/10.1103/PhysRevX.5.041025} {\bibfield  {journal} {\bibinfo  {journal} {Phys. Rev. X}\ }\textbf {\bibinfo {volume} {5}},\ \bibinfo {pages} {041025} (\bibinfo {year} {2015})}\BibitemShut {NoStop}%
\bibitem [{\citenamefont {Anandan}\ and\ \citenamefont {Aharonov}(1990)}]{anandan1990geometry}%
  \BibitemOpen
  \bibfield  {author} {\bibinfo {author} {\bibfnamefont {J.}~\bibnamefont {Anandan}}\ and\ \bibinfo {author} {\bibfnamefont {Y.}~\bibnamefont {Aharonov}},\ }\href {https://doi.org/10.1103/PhysRevLett.65.1697} {\bibfield  {journal} {\bibinfo  {journal} {Phys. Rev. Lett.}\ }\textbf {\bibinfo {volume} {65}},\ \bibinfo {pages} {1697} (\bibinfo {year} {1990})}\BibitemShut {NoStop}%
\bibitem [{sm()}]{sm}%
  \BibitemOpen
  \href@noop {} {}\bibinfo {note} {For details of the calculations, simulations, proofs and miscellaneous information, see Supplemental Material.}\BibitemShut {Stop}%
\bibitem [{\citenamefont {Chen}\ and\ \citenamefont {Muga}(2010)}]{chen2010transient}%
  \BibitemOpen
  \bibfield  {author} {\bibinfo {author} {\bibfnamefont {X.}~\bibnamefont {Chen}}\ and\ \bibinfo {author} {\bibfnamefont {J.~G.}\ \bibnamefont {Muga}},\ }\href {https://doi.org/10.1103/PhysRevA.82.053403} {\bibfield  {journal} {\bibinfo  {journal} {Phys. Rev. A}\ }\textbf {\bibinfo {volume} {82}},\ \bibinfo {pages} {053403} (\bibinfo {year} {2010})}\BibitemShut {NoStop}%
\bibitem [{\citenamefont {Deffner}\ and\ \citenamefont {Lutz}(2013)}]{deffner2013energytime}%
  \BibitemOpen
  \bibfield  {author} {\bibinfo {author} {\bibfnamefont {S.}~\bibnamefont {Deffner}}\ and\ \bibinfo {author} {\bibfnamefont {E.}~\bibnamefont {Lutz}},\ }\href {https://doi.org/10.1088/1751-8113/46/33/335302} {\bibfield  {journal} {\bibinfo  {journal} {Journal of Physics A: Mathematical and Theoretical}\ }\textbf {\bibinfo {volume} {46}},\ \bibinfo {pages} {335302} (\bibinfo {year} {2013})}\BibitemShut {NoStop}%
\bibitem [{\citenamefont {Mohan}\ and\ \citenamefont {Pati}(2021)}]{mohan2021reverse}%
  \BibitemOpen
  \bibfield  {author} {\bibinfo {author} {\bibfnamefont {B.}~\bibnamefont {Mohan}}\ and\ \bibinfo {author} {\bibfnamefont {A.~K.}\ \bibnamefont {Pati}},\ }\href {https://doi.org/10.1103/PhysRevA.104.042209} {\bibfield  {journal} {\bibinfo  {journal} {Phys. Rev. A}\ }\textbf {\bibinfo {volume} {104}},\ \bibinfo {pages} {042209} (\bibinfo {year} {2021})}\BibitemShut {NoStop}%
\bibitem [{\citenamefont {Hellinger}(1909)}]{hellinger1909neue}%
  \BibitemOpen
  \bibfield  {author} {\bibinfo {author} {\bibfnamefont {E.}~\bibnamefont {Hellinger}},\ }\href {https://doi.org/10.1515/crll.1909.136.210} {\bibfield  {journal} {\bibinfo  {journal} {Journal für die reine und angewandte Mathematik}\ }\textbf {\bibinfo {volume} {1909}},\ \bibinfo {pages} {210–271} (\bibinfo {year} {1909})}\BibitemShut {NoStop}%
\bibitem [{\citenamefont {Kim}\ \emph {et~al.}(2022)\citenamefont {Kim}, \citenamefont {Murugan}, \citenamefont {Olle},\ and\ \citenamefont {Rosa}}]{kim2022operator}%
  \BibitemOpen
  \bibfield  {author} {\bibinfo {author} {\bibfnamefont {J.}~\bibnamefont {Kim}}, \bibinfo {author} {\bibfnamefont {J.}~\bibnamefont {Murugan}}, \bibinfo {author} {\bibfnamefont {J.}~\bibnamefont {Olle}},\ and\ \bibinfo {author} {\bibfnamefont {D.}~\bibnamefont {Rosa}},\ }\href {https://doi.org/10.1103/PhysRevA.105.L010201} {\bibfield  {journal} {\bibinfo  {journal} {Phys. Rev. A}\ }\textbf {\bibinfo {volume} {105}},\ \bibinfo {pages} {L010201} (\bibinfo {year} {2022})}\BibitemShut {NoStop}%
\bibitem [{\citenamefont {Divi}\ \emph {et~al.}(2025)\citenamefont {Divi}, \citenamefont {Murugan},\ and\ \citenamefont {Rosa}}]{divi2024sykchargingadvantagerandom}%
  \BibitemOpen
  \bibfield  {author} {\bibinfo {author} {\bibfnamefont {F.}~\bibnamefont {Divi}}, \bibinfo {author} {\bibfnamefont {J.}~\bibnamefont {Murugan}},\ and\ \bibinfo {author} {\bibfnamefont {D.}~\bibnamefont {Rosa}},\ }\href {https://doi.org/10.1103/PhysRevB.111.075138} {\bibfield  {journal} {\bibinfo  {journal} {Phys. Rev. B}\ }\textbf {\bibinfo {volume} {111}},\ \bibinfo {pages} {075138} (\bibinfo {year} {2025})}\BibitemShut {NoStop}%
\end{thebibliography}%


\begin{thebibliography}{3}%
\makeatletter
\providecommand \@ifxundefined [1]{%
 \@ifx{#1\undefined}
}%
\providecommand \@ifnum [1]{%
 \ifnum #1\expandafter \@firstoftwo
 \else \expandafter \@secondoftwo
 \fi
}%
\providecommand \@ifx [1]{%
 \ifx #1\expandafter \@firstoftwo
 \else \expandafter \@secondoftwo
 \fi
}%
\providecommand \natexlab [1]{#1}%
\providecommand \enquote  [1]{``#1''}%
\providecommand \bibnamefont  [1]{#1}%
\providecommand \bibfnamefont [1]{#1}%
\providecommand \citenamefont [1]{#1}%
\providecommand \href@noop [0]{\@secondoftwo}%
\providecommand \href [0]{\begingroup \@sanitize@url \@href}%
\providecommand \@href[1]{\@@startlink{#1}\@@href}%
\providecommand \@@href[1]{\endgroup#1\@@endlink}%
\providecommand \@sanitize@url [0]{\catcode `\\12\catcode `\$12\catcode `\&12\catcode `\#12\catcode `\^12\catcode `\_12\catcode `\%12\relax}%
\providecommand \@@startlink[1]{}%
\providecommand \@@endlink[0]{}%
\providecommand \url  [0]{\begingroup\@sanitize@url \@url }%
\providecommand \@url [1]{\endgroup\@href {#1}{\urlprefix }}%
\providecommand \urlprefix  [0]{URL }%
\providecommand \Eprint [0]{\href }%
\providecommand \doibase [0]{https://doi.org/}%
\providecommand \selectlanguage [0]{\@gobble}%
\providecommand \bibinfo  [0]{\@secondoftwo}%
\providecommand \bibfield  [0]{\@secondoftwo}%
\providecommand \translation [1]{[#1]}%
\providecommand \BibitemOpen [0]{}%
\providecommand \bibitemStop [0]{}%
\providecommand \bibitemNoStop [0]{.\EOS\space}%
\providecommand \EOS [0]{\spacefactor3000\relax}%
\providecommand \BibitemShut  [1]{\csname bibitem#1\endcsname}%
\let\auto@bib@innerbib\@empty
\bibitem [{\citenamefont {Rossini}\ \emph {et~al.}(2020)\citenamefont {Rossini}, \citenamefont {Andolina}, \citenamefont {Rosa}, \citenamefont {Carrega},\ and\ \citenamefont {Polini}}]{rossini2020quantum}%
  \BibitemOpen
  \bibfield  {author} {\bibinfo {author} {\bibfnamefont {D.}~\bibnamefont {Rossini}}, \bibinfo {author} {\bibfnamefont {G.~M.}\ \bibnamefont {Andolina}}, \bibinfo {author} {\bibfnamefont {D.}~\bibnamefont {Rosa}}, \bibinfo {author} {\bibfnamefont {M.}~\bibnamefont {Carrega}},\ and\ \bibinfo {author} {\bibfnamefont {M.}~\bibnamefont {Polini}},\ }\href {https://doi.org/10.1103/PhysRevLett.125.236402} {\bibfield  {journal} {\bibinfo  {journal} {Phys. Rev. Lett.}\ }\textbf {\bibinfo {volume} {125}},\ \bibinfo {pages} {236402} (\bibinfo {year} {2020})}\BibitemShut {NoStop}%
\bibitem [{\citenamefont {V.~Romero}\ and\ \citenamefont {Santos-Suárez}(2023)}]{paulicomposer2023}%
  \BibitemOpen
  \bibfield  {author} {\bibinfo {author} {\bibfnamefont {S.}~\bibnamefont {V.~Romero}}\ and\ \bibinfo {author} {\bibfnamefont {J.}~\bibnamefont {Santos-Suárez}},\ }\href {https://doi.org/10.1007/s11128-023-04204-w} {\bibfield  {journal} {\bibinfo  {journal} {Quantum Information Processing}\ }\textbf {\bibinfo {volume} {22}},\ \bibinfo {pages} {449} (\bibinfo {year} {2023})}\BibitemShut {NoStop}%
\bibitem [{\citenamefont {Al-Mohy}\ and\ \citenamefont {Higham}(2011)}]{almohy2011computing}%
  \BibitemOpen
  \bibfield  {author} {\bibinfo {author} {\bibfnamefont {A.~H.}\ \bibnamefont {Al-Mohy}}\ and\ \bibinfo {author} {\bibfnamefont {N.~J.}\ \bibnamefont {Higham}},\ }\href {https://doi.org/10.1137/100788860} {\bibfield  {journal} {\bibinfo  {journal} {SIAM Journal on Scientific Computing}\ }\textbf {\bibinfo {volume} {33}},\ \bibinfo {pages} {488} (\bibinfo {year} {2011})}\BibitemShut {NoStop}%
\end{thebibliography}%
\clearpage

\end{document}


\title{Supplemental Material for:\\``Scrambling in the Charging of Quantum Batteries''}
\author{Sebastián V. Romero$^{\orcidlink{0000-0002-4675-4452}}$}
\affiliation{Department of Physical Chemistry, University of the Basque Country UPV/EHU, Apartado 644, 48080 Bilbao, Spain}
\affiliation{Kipu Quantum GmbH, Greifswalderstrasse 212, 10405 Berlin, Germany}
\author{Yongcheng Ding$^{\orcidlink{0000-0002-6008-0001}}$}
\affiliation{Department of Physical Chemistry, University of the Basque Country UPV/EHU, Apartado 644, 48080 Bilbao, Spain}
\affiliation{Institute for Quantum Science and Technology, Department of Physics, Shanghai University, Shanghai 200444, China}
\author{Xi Chen$^{\orcidlink{0000-0003-4221-4288}}$}
\email{xi.chen@csic.es}
\affiliation{Instituto de Ciencia de Materiales de Madrid (CSIC), Cantoblanco, E-28049 Madrid, Spain}
\author{Yue Ban$^{\orcidlink{0000-0003-1764-4470}}$}
\email{yue.ban@csic.es}
\date{\today}

\begin{abstract}
    In this Supplemental Material further notes and extended results are provided to support and elucidate the findings attached in the main text. In particular, a thorough study of the energy variances involved in the charging protocol and their corresponding system size dependencies are presented, being the sources of the characteristic superextensive scaling of complex SYK quantum batteries. Additionally, we study in detail the spectrum properties of our battery Hamiltonian, which can be used to analyze quantum battery performance and the effect of scrambling over an initial state. Finally, a detailed sparsity analysis of our system is presented to motivate the numerical methods used in our work to reduce the computational resources needed for our simulations.
\end{abstract}

\maketitle

\renewcommand{\thetable}{S\arabic{table}}
\renewcommand{\theequation}{S\arabic{equation}}
\renewcommand{\thefigure}{S\arabic{figure}}
\renewcommand{\bibnumfmt}[1]{[S#1]}
\renewcommand{\citenumfont}[1]{S#1}

\section*{Analysis of energy variances and charger bandwidths}\label{app:variances}

A key part of the results obtained in our study relies on the dynamics of the energy variances and their system size dependence, building blocks of the power bounds and quantum speed limits computed in our work as well as to thoroughly unveil the quantum origin of the quantum advantage granted by complex SYK quantum batteries. Since the battery Hamiltonian $\mathcal{H}_0$ is local, it is possible to split easily its variance into their local ($\Delta^\text{loc}_\tau \mathcal{H}_0^2$) and entangled ($\Delta^\text{ent}_\tau \mathcal{H}_0^2$) contributions as
\begin{equation}\label{eq:locent_h0}
\begin{aligned}
 \Delta^\text{loc}_\tau \mathcal{H}_0^2 &= \frac{1}{\tau}\int_0^\tau \text{d}t \sum_{j=1}^N\left[\braket{h_j^2}_t-\braket{h_j}_t^2\right], \\
 \Delta^\text{ent}_\tau \mathcal{H}_0^2 &= \frac{1}{\tau}\int_0^\tau \text{d}t\sum_{i\neq j}\left[\braket{h_ih_j}_t-\braket{h_i}_t\!\braket{h_j}_t\right],
\end{aligned}%
\end{equation}%
with $\Delta_\tau \mathcal{H}_0^2 = \Delta^\text{loc}_\tau \mathcal{H}_0^2 + \Delta^\text{ent}_\tau \mathcal{H}_0^2$ (in units of $\omega_0^2$). While the local contribution is extensive by construction, whose sum of local terms scales linearly with system size, the entangled contribution accounts the correlations between different quantum cells, source of a potential superlinear scaling with system size, thus a potential source of quantum advantage on the charging of quantum batteries~\cite{rossini2020quantum}.
\begin{figure*}[!tb]
    \centering
    \includegraphics[width=.5\linewidth]{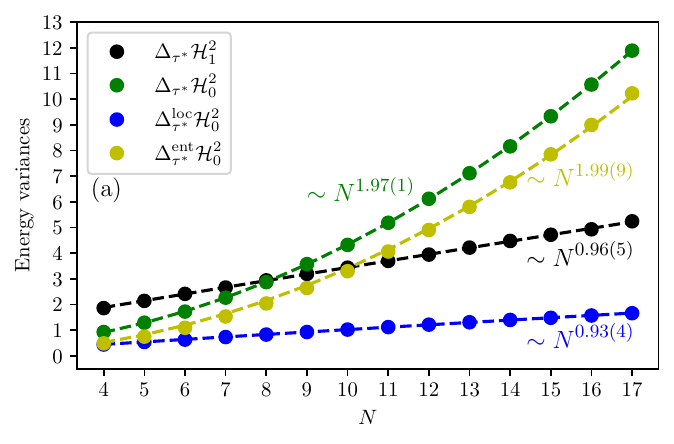}%
    \includegraphics[width=.5\linewidth]{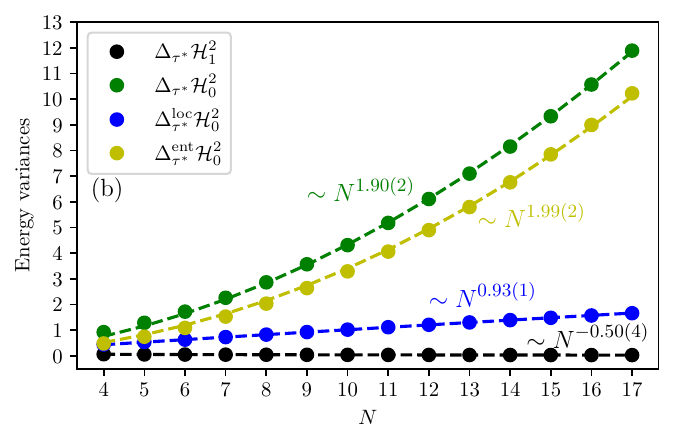}\vspace{-2mm}%
    \caption{(a) Variances of the battery and nonregularized charger Hamiltonians (in units of $\omega_0^2$ and $J^2$, respectively), splitting the variances of the battery into their local and entangled contributions following~\eqref{eq:locent_h0}. (b) Variances of the battery and regularized charger Hamiltonians. Data is fitted to a power-law function $a+bN^c$ (dashed lines) with their corresponding scalings marked, where the three data points corresponding to the smallest $N$ have been discarded from the fits.}\label{fig:fig7}
\end{figure*}%

In~\figlabel{fig:fig7} we compute these variances for the battery and charger Hamiltonians for both nonregularized and regularized frameworks, supporting our previous discussions and findings on how they scale separately. As expected, $\Delta^\text{loc}_\tau \mathcal{H}_0^2$ displays an extensive scaling but, contrarily, $\Delta^\text{ent}_\tau \mathcal{H}_0^2$ showcases a superextensive growth that permits a quadratic growth of the battery Hamiltonian variance with system size, as expected. Between both frameworks, the only variance that significantly changes comes from the charger Hamiltonian where, as discussed in the main text, this fact naturally comes because of removing energy units in this bare representation, being both representations interchangeably just by multiplying variances with bandwidths accordingly. In~\figlabel{fig:fig8} the charging Hamiltonian bandwidth scaling with system size $N$ is shown. Since the bandwidth of the SYK model scales as $\mathcal{O}[N]$ in the large $N$ limit, the slight mismatch in our results is related to finite size effects, which become negligible for larger system sizes.
\begin{figure}[!tb]
    \centering
    \includegraphics[width=.55\linewidth]{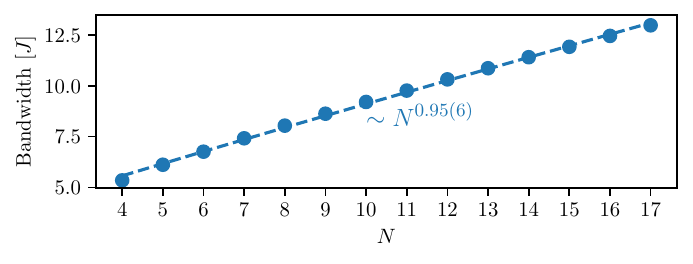}\vspace{-2mm}%
    \caption{Bandwidth scaling of the charging Hamiltonian (in units of $J$) simulated in our work against system size. Data is fitted to a power-law function $a+bN^c$ (dashed line), where the three data points corresponding to the smallest $N$ have been discarded from the fits, with their corresponding system size scalings annotated.}\label{fig:fig8}
\end{figure}%

\section*{Exact diagonalization of the local battery}\label{app:times}

For the sake of clarity, here we attach additional information and results to enlighten the nature of our chosen local battery Hamiltonian and how their eigenstates evolve when the charging Hamiltonian is introduced. The spectrum of $\mathcal{H}_0$ splits into multiples of $\omega_0$, letting a simple but thorough study of how the energy is stored along the evolution and how the initial state scrambles. This fact can be easily proven diagonalizing the battery Hamiltonian $\mathcal{H}_0$, where
\begin{equation}\label{eq:h0_diag}
  \mathcal{H}_0=\frac{\omega_0}{2}\sum_{j=1}^N\sigma^y_j=\frac{\omega_0}{2}\sum_{j=1}^N[SD(-1,1)S^{-1}]_j =\frac{\omega_0}{2}\Bigg(\bigotimes_{j=1}^NS_j\Bigg)\sum_{j=1}^ND_j(-1,1)\Bigg(\bigotimes_{j=1}^NS^{-1}_j\Bigg),
\end{equation}
with $S\coloneqq \begin{bmatrix}\ket{\downarrow^y} & \ket{\uparrow^y}\end{bmatrix}$ (where $\sigma^y\ket{\updownarrows^y}=\pm\ket{\updownarrows^y}$) and $D(-1,1)=\text{diag}(-1,1)$. The spectrum of $\mathcal{H}_0$ is given by $\epsilon_k=(2k-N)\omega_0/2$ (multiplicity $\binom{N}{k}$) $\forall k\in[0,N]$, whose ground state is $\ket{\psi(0)}\coloneqq\bigotimes_{j=1}^N\ket{\downarrow^y}_j$ with $\epsilon_0=-N\omega_0/2$. In our work, we shift the energies adding the ground state contribution, thus $\epsilon_k\mapsto\epsilon_k+N\omega_0/2$, returning $\epsilon_k=k\omega_0$ as expected. This particular binomial-distributed degeneracy of the spectrum, split into $N+1$ energy levels, favours a simplistic study of how the probability of occupying eigenstates carrying an amount of energy $\epsilon_k$ evolves. Therefore, it allows to have a clear picture of how the energy is stored along time. For this latter purpose, let $\ket{\psi(\tau)}=e^{-i\mathcal{H}_1\tau}\ket{\psi(0)}$ be the state after charging completion and decompose the battery Hamiltonian as $\mathcal{H}_0=\sum_{k=0}^N\epsilon_k\sum_i\proj{k,i}$ with $\epsilon_k= k\omega_0$ its eigenvalues after shifting ground state and $\ket{k,i}$ the $i$th degenerate eigenstate corresponding to $\epsilon_k$. With the system initialized at $\mathcal{H}_0$ ground state, the population dynamics evolve as $p_k(\tau)\coloneqq\sum_i|\braket{k,i|\psi(\tau)}|^2$. 

\section*{Exploiting sparsity and numerical methods}

In the next lines, a sparsity analysis of the Hamiltonians involved in our formulation is shown. The main bottleneck of our calculations falls on computing the Hamiltonians considered in our system and, in particular, the time-evoution operators $e^{-i\mathcal{H}_1t}$. In our case, exploiting sparse matrix techniques can drastically save memory resources and reduce computational costs.

First, the number of nonzero entries of the battery Hamiltonian $\mathcal{H}_0$ is $N2^N$~\cite{paulicomposer2023}, with $N$ the number of quantum cells considered. For the complex SYK charging Hamiltonian $\mathcal{H}_1$, in~\tablabel{tab:nnz} a discussion on the number of nonzero entries is attached, which scales as $\mathcal{O}[N^42^N]$. The maximum matrix density in the charging protocol is acquired when the switching function satisfies $\lambda(t)\in(0,1)$, whose number of nonzeros is directly given by the sum of both $\mathcal{H}_0$ and $\mathcal{H}_1$ Hamiltonians separately.

From here, we want to efficiently tackle the time-evolved state $\ket{\psi(t)}=e^{-i\mathcal{H}_1t}\ket{\psi(0)}$ computation at different times within the charging time window $t\in[0,\tau]$, exploiting the large sparsity of $\mathcal{H}_1$. It is possible to considerably leverage its calculation by applying the algorithm presented in Ref.~\cite{almohy2011computing}, whose method computes efficiently $e^{t_kA}B$ with $t_k$ within an equally spaced grid of points, $A$ is an $n\times n$ matrix and $B$ an $n\times n_0$ matrix with $n_0\ll n$. This constraint is met for our case, since $n=2^N$ and $n_0=1$.

For numerical integration of the variances, we use the composite Simpson's 1/3 rule, where the integral between $[0,\tau]$ split in $n$ steps has an associated bounded approximation error proportional to $\tau^5/n^4$.
\begin{table}[!tb]
\centering
\begin{tabular}{|r|r|r|}\hline
    \multicolumn{1}{|c|}{Interaction term} & \multicolumn{1}{c|}{Number of nonzeros} & \multicolumn{1}{c|}{Order} \\ \hline
    $c^\dagger_i c^\dagger_j c_k c_l$ & $2^{N-4}\binom{N}{2}\binom{N-2}{2}$ & $\mathcal{O}[N^42^N]$ \\
    $c^\dagger_i n_j c_l$ & $2(2^{N-2}-1)\binom{N}{2}$ & $\mathcal{O}[N^22^N]$ \\
    $n_in_j$ & $2^N-N-1$ & $\mathcal{O}[2^N]$ \\ \hline
\end{tabular}\vspace{3mm}
\caption{Number of nonzero elements per interaction for the complex SYK Hamiltonian for $N$ quantum cells. The density decays as $\mathcal{O}[N^42^{-N}]$, encouraging a sparse treatment.}\label{tab:nnz}
\end{table}%

\bibliography{bibfile}
\clearpage